\documentclass[a4paper]{article}

\usepackage{INTERSPEECH2020}

\title{Learning Speaker Representation with Semi-supervised Learning approach for Speaker Profiling}
\name{Shangeth Rajaa$^1$, Pham Van Tung$^2$, Chng Eng Siong$^2$}
\address{
  $^1$skit.ai, India\\
  $^2$School of Computer Science and Engineering, Nanyang Technological University, Singapore}
\email{shangeth.rajaa@skit.ai, vtpham@ntu.edu.sg, ASESChng@ntu.edu.sg}
\usepackage{graphicx}
\graphicspath{ {./imgs/} }

\usepackage{multirow}
\usepackage{colortbl}
\begin{document}

\maketitle
\begin{abstract} 
Speaker profiling, which aims to estimate speaker characteristics such as age and height, has a wide range of applications in forensics, recommendation systems, etc. In this work, we propose a semi-supervised learning approach to mitigate the issue of low training data for speaker profiling. This is done by utilizing external corpus with speaker information to train a better representation which can help to improve the speaker profiling systems. Specifically, besides the standard supervised learning path, the proposed framework has two more paths: (1) an unsupervised speaker representation learning path that helps to capture the speaker information; (2) a consistency training path that helps to improve the robustness of the system by enforcing it to produce similar predictions for utterances of the same speaker. The proposed approach is evaluated on the TIMIT and NISP datasets for age, height, and gender estimation, while the Librispeech is used as the unsupervised external corpus. Trained both on single-task and multi-task settings, our approach was able to achieve state-of-the-art results on age estimation on the TIMIT Test dataset with Root Mean Square Error(RMSE) of 6.8 and 7.4 years and Mean Absolute Error(MAE) of 4.8 and 5.0 years for male and female speakers respectively. 
\end{abstract}


\section{Introduction}

Speech is a basic and very important mode of communication that can convey the content that the speaker wants to communicate. The speech signal also contains information about the speaker's origin, gender, emotion, and identity of the speaker. With speaker profiling systems we try to estimate the speaker's age, height, and gender with the speech signal. These systems have a wide range of applications in forensics, recommender systems, and many biometric applications to identify the speaker \cite{schilling_marsters_2015} \cite{1262027}.

\textbf{Related Works : } Research \cite{Laver1979PhoneticAL} has shown that the build of a person can affect speech production and there is a positive correlation between the vocal tract length of a person and their height \cite{doi:10.1121/1.1835958}. The age and gender of the speaker can affect some voice characteristics like fundamental frequency and speech rate \cite{LI2013151}. 

Most of the speech predictive tasks involve extracting the essential features from the raw speech signal and use the extracted features with a Machine Learning model for prediction. Multiple features are proposed and used for the speaker profiling task in the past like sub-glottal resonance, fundamental frequency, statistical and spectral features of the signal, etc. \cite{Dusan2005EstimationOS} estimates the height and the vocal tract length of the speaker and studies the correlation of features like Mel-frequency Cepstrum Coefficients (MFCC), formant frequencies, fundamental frequency and Linear Predictive Coding(LPC). Few methods \cite{muller2006automatic} \cite{LI2013151} \cite{mallouh2018new} try to classify the age and height group to which the speaker belongs using the speech features with machine learning models such as Support Vector Machines(SVM), Artificial Neural Networks(ANN), etc. Short-term temporal features at multiple resolutions are used as speech representation in \cite{7449696} to estimate the height and age of the speaker.

In recent years Deep Neural Networks(DNN) models are extensively used to achieve the state of the art result in multiple speech tasks including speaker profiling \cite{8683397}. Regression models like SVR, ANN are used with i-vectors \cite{7296469} and x-vectors \cite{Ghahremani2018} of the speech signals for estimating the height of the speaker. Long Short Term Memory(LSTM) models are also used with short speech utterances for estimating height, age \cite{zazo2018age} and gender \cite{ertam2019effective} of the speaker. Features are directly extracted from the raw speech signals using a Convolutional Neural network(CNN) which are then used for classification \cite{ertam2019effective} or regression tasks. \cite{kwasny2020joint} uses a CNN-based model which is pretrained on VoxCeleb and Common Voice dataset and finetuned on TIMIT dataset \cite{garofolo1993darpa} to estimate the age and gender of the speaker in a multi-task setting and achieves an accuracy of $99.6\%$ on gender classification.

Many previous methods required a long-duration speech signal to estimate the height, age, gender of a speaker and some work needs the phoneme information \cite{Dusan2005EstimationOS} of the speech utterance. Lack of supervised training dataset is also a factor that affects the performance of the speaker profiling systems. Semi-supervised learning \cite{zhu2009introduction} approaches can aid in solving the problem of lack of supervised data, by learning the general representation with the help of an unsupervised dataset and combining them with the supervised dataset to learn the downstream supervised task.  Semi-supervised learning/Unsupervised learning has been used in multiple works \cite{sholokhov2018semi} \cite{schneider2019wav2vec} \cite{kawakami2020learning} \cite{song2019speech} to learn the representation of the speech signal for some downstream tasks like speech recognition, speaker recognition, etc.
The major drawback of using the previous semi-supervised/unsupervised approach for speaker profiling is the learned representations may contain useless features like the content of the speech, which may not help the speaker profiling task. Speaker profiling task may require the speaker's information \cite{ravanelli2018learning}, so the representation should be rich in speaker information and ignore other unnecessary information in the speech signal.


\textbf{Our Contributions : }In this work we attempt to improve the performance of the speaker profiling task with a semi-supervised learning approach. Previous approaches train supervised models for speaker profiling, we try to use the speaker information to guide the speaker profiling task. We train an encoder to produce a high-level speaker representation by maximizing the mutual information \cite{hjelm2018learning} of the representations of speech signals of similar speakers. Most of the previous works focus on estimating the height/age/gender of the speaker separately, we attempt to estimate all three parameters of the speaker in a multi-task setting with a short length speech signal of 4 seconds of a raw audio speech signal and compare it with single-task models which estimate only height/age/gender. Previous works also train different models for males and females, we train a single model for both genders. The experiment results show that our approach learns useful speaker representations from raw speech signals which also gives better results in the speaker profiling task than the previous approaches.

The paper is organized as follows. In Section 2 we describe the semi-supervised training method for speaker profiling. Section 3 describes the dataset used for the experiments, proposed neural network architectures, and the training setup. Section 4 gives the results of the experiments for height, age, and gender estimation and previous results. Finally, the paper is concluded in Section 6.


\section{Method}
\begin{figure}[h]
\centering
\includegraphics[width=\linewidth]{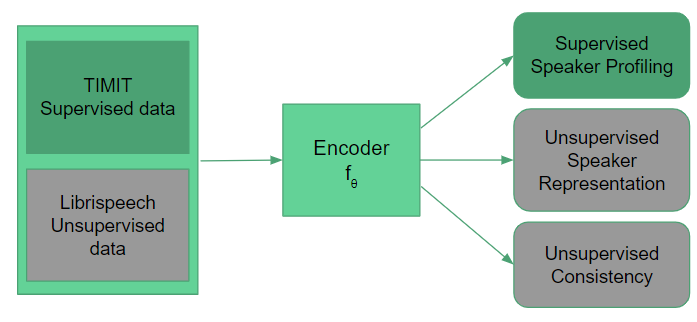}
\caption{The proposed semi-supervised learning framework. All 3 tasks share the same encoder. Unsupervised speaker representation helps the encoder to capture better speaker information. Unsupervised Consistency helps to improve the robustness of the model by forcing it to produce similar predictions for utterances belonging to the same speaker.}
\label{fig:1}
\end{figure}

Our approach is a semi-supervised learning method that has three paths as shown in Figure \ref{fig:1}: (1) The supervised path which estimates the height, age, and gender of the speaker; (2) the unsupervised representation learning path which uses an  unsupervised\footnote{We note that although this dataset has speaker labels, it does not have height, age information. As such, we consider it an unsupervised data} external dataset to learn the speaker representation; (3) the consistency path which makes the predictions robust. The speaker representation is learned by maximizing the mutual information between the representation of speech signals of the same speaker unlike \cite{ravanelli2018learning} which tries to maximize the mutual information between chunks of speech sampled from the same sentence/utterance. The motivation behind the unsupervised speaker representation learning is that the speaker representation should contain features that can improve the speaker profiling task better than raw audio signal which may contain useless information for the speaker profiling task such as the phonemes, background information, etc. 

\subsection{Supervised Speaker Profiling}
The supervised path estimates the height, age, and gender of the speaker from the raw audio signal. This path has two networks, the encoder and the regressor(regression for height, age, and classification for gender). The encoder network $f_{\theta}: \mathbb{R}^K \rightarrow \mathbb{R}^N$ transforms the speech signal $X$ of $K$ dimension into a $N$ dimensional latent code $z = f_{\theta}(X)$. The regressor network $h_{\psi}: \mathbb{R}^N \rightarrow \mathbb{R}^3$, takes the latent code $z$ of the speech signal as input and estimates the height, age, and gender of the speaker. $\hat{y}_h, \hat{y}_a, \hat{y}_g = h_{\psi}(z)$, where $\hat{y}_h, \hat{y}_a, \hat{y}_g$ are the estimated height, age and gender respectively. This supervised path is trained in a multi-task setting with the loss function as

\begin{equation}
L_{p}(\theta, \phi) = \alpha L_{reg}(\hat{y}_h, y_h) + \beta L_{reg}(\hat{y}_a, y_a) + \gamma L_{cls}(\hat{y}_g, y_g)
\label{eq:1}
\end{equation}

where $L_{reg}$, $L_{cls}$ are the regression and classification loss function, $y_h, y_a, y_g$ are the target value for height, age and gender respectively from the supervised dataset. For single-task model the loss function is $ L_{reg}$ or $L_{cls}$ depending on the task.

\begin{figure}[h]
\centering
\includegraphics[width=0.9\linewidth]{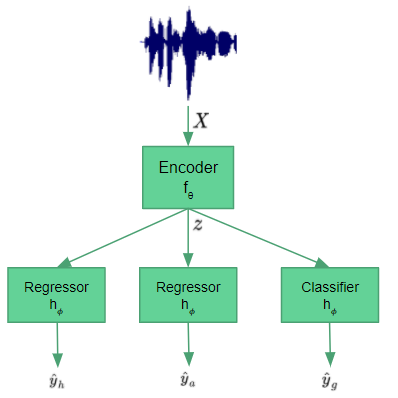}
\caption{Supervised Speaker Profiling path}
\end{figure}

We use a CNN-LSTM model for the encoder $f_{\theta}$ with 1-dimensional convolutional layers and the output of the final timestep of the LSTM layer is the encoded representation $z$. The regressor $h_{\psi}$ is a 2 layer ANN network. 

\subsection{Unsupervised Speaker Representation learning}
The unsupervised learning path aims to learn the speaker representation $z = f_{\theta}(X)$ which can best represent the speaker information from the raw audio speech signal $X$ by training the encoder network $f_{\theta}$. The discriminator network $g_{\omega }: \mathbb{R}^{2N} \rightarrow \mathbb{R}$ is trained to distinguish between the latent vector pairs of same speakers $(X, X_p)$ and different speakers $(X, Xn)$, where $X$ is the speech signal of the anchor speaker, $X_p$ is the positive speech signal which is sampled from the same speaker as $X$ and $X_n$ is the negative sample which is sampled from a different speaker than $X$. 

The only common information between the anchor signal $X$ and the positive signal $X_p$ is the speaker information. The encoder should learn to capture the speaker information from the speech signals which can be used by the discriminator to distinguish between positive and negative pairs. This learning process enables the encoder to disentangle the speaker information from other irrelevant information in the speech signal like the environment, phonemes, etc. 

The results from  \cite{ravanelli2018learning} show that using a Mutual information based loss function outperforms distance-based loss functions like Triplet Loss \cite{schroff2015facenet}. In, mutual information based approaches, Binary Cross-Entropy(BCE) Loss as a loss function to distinguish between the speech pairs learns better speaker representation compared to other methods like Mutual Information Neural Estimation(MINE) \cite{belghazi2018mutual} and Noise
Contrastive Estimation(NCE) \cite{oord2018representation} for speaker recognition task. So BCE loss is used as the loss function to learn the speaker representations.

\begin{equation}
L_{repr}(\theta, \omega )=\mathbb{E}_{X_{p}}\left[\log \left(g\left(z, z_{p}\right)\right)\right]+\mathbb{E}_{X_{n}}\left[\log \left(1-g\left(z, z_{n}\right)\right)\right]
\label{eq:2}
\end{equation}

where $\mathbb{E}_{X_{p}}$ is the expectation over positive samples and $\mathbb{E}_{X_{n}}$ is the expectation over negative samples and $z, z_p, z_n$ are the encoded representations of $X, X_p, X_n$ respectively. This estimates the Jensen-Shannon divergence between the distributions and not the exact KL divergence in the definition of Mutual Information.

Each sample of the unsupervised dataset will have 3 speech signals $(X, X_p, X_n)$. At each iteration, the anchor speaker is chosen randomly and the anchor $X$ and positive utterance $X_p$ are sampled from the anchor speaker, and the negative utterance $X_n$ is sampled from a different speaker randomly. The discriminator $g_{\omega}$ is also a 2 layer ANN model.

\vspace{-1em}
\begin{figure}[h]
\centering
\includegraphics[width=0.9\linewidth]{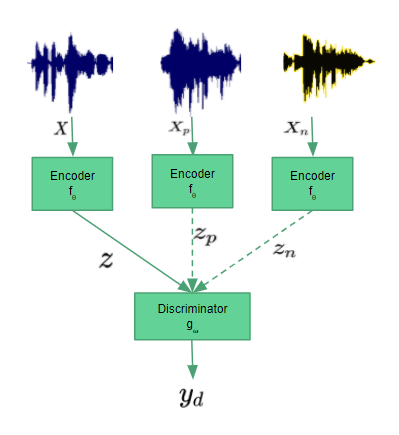}
\caption{Unsupervised Speaker Representation path}
\end{figure}

\subsection{Unsupervised Consistency Training}
Along with the speaker representation learning path, we also perform consistency training on the speaker profiling task with the unsupervised dataset. We train the encoder, regressor to reduce the distance between the estimated height, age, and gender of $X$ and $X_p$ which is sampled from the same speaker in the unsupervised dataset. These speech signal does not have the supervised labels for height, age, and gender, but as the signals are from the same speakers, they should have the same height, age and gender values. The loss function used in the consistency loss is the same as the profiling loss $L_p$.

\begin{equation}
L_{c}(\theta, \phi) = \alpha L_{reg}(\hat{y}_h, \hat{y}_{ph}) + \beta L_{reg}(\hat{y}_a, \hat{y}_{pa}) + \gamma L_{cls}(\hat{y}_g, \hat{y}_{pg})
\label{eq:3}
\end{equation}

Where $\hat{y}_h, \hat{y}_a, \hat{y}_g$ and $\hat{y}_{ph}, \hat{y}_{pa}, \hat{y}_{pg}$ are the predicted age, height, gender of $X$ and  $X_p$ respectively. This helps to enforce the height, age, and gender of the unlabelled speech signals of the same speaker to be similar. Augmentation methods are applied to speech signals which are discussed in the experimental setup section. This will act as a regularization for the speaker profiling regressor network and help to generalize better. This technique has improved the performance of multiple supervised tasks in the image/text domain \cite{xie2019unsupervised}.

\begin{figure}[h]
\centering
\includegraphics[width=\linewidth]{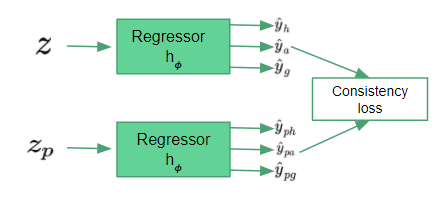}
\caption{Unsupervised Consistecy Training}
\end{figure}

When the speaker profiling system is train only wi the supervised path, the model tends to overfit due to lack of data, so we use the unsupervised path and consistency path to regularize the training.
The Encoder $f_{\theta}$, regressor $h_{\phi}$ and discriminator $g_{\omega}$ are trained in a multi-task setting with all the three tasks discussed above. The final loss function for training the model is given by

\begin{equation}
L(\theta, \phi, \omega) = L_{p} + L_{repr} + L_{c} 
\label{eq:4}
\end{equation}

\section{Experimental Setup}
\subsection{Dataset}
This paper uses the TIMIT corpus as the supervised dataset to estimate height, age, gender. TIMIT dataset has a total of 630 speakers with 461 speakers(135 female and 326 male) for the training set and 168 speakers(56 female and 112 male) for the test set. The range of age for the training set is 21 years to 76 years and 22 years to 68 years for the test set. The range of height for the training set is 145cm to 199 cm and 153 cm to 204 cm for the test set. The average length of the speech recording in the dataset is about 2.5 seconds. We also train our approach on the NISP \cite{kalluri2020nisp} dataset, which is a multi-lingual corpus with 5 different Indian languages along with English. It also has the speaker's information such as height, age, weight, shoulder width. Training on the NISP train dataset is done in a multi-task setting to estimate the height, age, gender, and the trained model is tested on the NISP test dataset.

We use LibriSpeech corpus \cite{panayotov2015librispeech} as the unsupervised speech corpus. It contains 960 hours of clean and noisy speech utterances. In this paper, we use the 360 hours clean dataset for training the unsupervised speaker representation and the consistency training paths. 


For the supervised dataset, $15\%$ of the training set speakers are used for the development set with an equal ratio split of male and female speakers.

\subsection{Training and Model Architecture}
The waveform of both the supervised dataset(TIMIT) and the unsupervised dataset(Librispeech) are cropped or padded to a fixed length of 4 seconds with a sampling rate of 16KHz, random cropping/padding in case of the training set, and center cropping/padding in case of the test set. The training waveforms are also augmented by adding random environmental noise. The height and age labels are standardized with mean and variance from the training set.

The CNN of the encoder $f_{\theta}$ has a similar architecture to the feature extractor of the wav2vec \cite{schneider2019wav2vec} model. The encoder has 5 layer CNN with kernel sizes of $(10, 8, 4, 4, 4)$ and strides of $(5, 4, 2, 2, 2)$ with group normalization, ReLU nonlinearity, and 512 channels. The final output $z$ of the LSTM network is a vector of dimension $512$. This CNN-LSTM encoder architecture encodes speech signals of any length to a single vector $z$ of size 512.

The regressor $h_{\phi}$ network has $[512, 128]$ hidden layers with ReLU nonlinearity. The regression loss $L_{reg}$ used is Mean Squared Error(MSE) and the classification loss $L_{cls}$ is Binary Cross-Entropy(BCE). The value of $\alpha, \beta, \gamma$ are chosen to be $1, 1, 0.1$ respectively.

The discriminator $g_{\omega}$ takes pairs of latent codes and returns a classification score $y_d$ of $1$ for $(z, z_p)$ and $0$ for $(z, z_n)$. The latent codes of the two speech signals are concatenated to 1024 dimensions and fed as an input to the discriminator. The number of hidden units is $[1024, 128]$. BCE loss $L_{repr}$ is used to classify if the two speech signals belong to the same speaker.

The parameters of all the networks are trained together to reduce the final loss function $L(\theta, \phi, \omega)$ which is the sum of supervised loss $L_{p}$ , representation loss $L_{repr}$ and consistency loss $L_{c}$ as shown in Eq \ref{eq:4}. The model is trained with DiffGrad optimizer \cite{DBLP:journals/corr/abs-1909-11015} with learning rate of $1e^{-3}$.


During training, each mini-batch consists of both supervised and unsupervised data. We empirically found the best ratio between the number of unsupervised data and that of supervised data is 4. The model is trained for 200 epochs and the checkpoint with the best validation loss is saved and tested on the test set.

Since our model is trained specifically for speaker representation learning, we use a pretrained wav2vec feature extractor with LSTM and dense layers for the prediction of height, age, and gender as the internal baseline as the wav2vec model is trained unsupervised for speech representation learning and performs well on a variety of speech tasks.

\section{Results}
The above-mentioned method is trained in two settings, single-task training where we train different models to predict height, age, and gender, and multi-task setting where height, age, and gender are predicted in a single model. We compare our results with an internal baseline and previous results. The test set of TIMIT and NISP corpus is used to compare the results for height, age, and gender estimation with previous methods.

We use the RMSE and MAE to evaluate the prediction error of the height and age and accuracy score to evaluate the gender classification. Table \ref{table:1} shows the comparison (RMSE and MAE) of our results with the previous methods for height and age. Table \ref{table:2} shows the results of gender accuracy. Table \ref{table:3} shows the comparison of our results with the baseline of the NISP dataset. We were also able to achieve an accuracy score of $1.0$ for gender classification in the NISP Test set. From our experiments, we observe that the performance of the Multi-task setting gave better results than the single-task setting in height and age estimation.

\begin{table}[h!]
\centering
\begin{tabular}{|c|c|c|c|c|c|}
\hline
\multirow{2}{*}{Method} & \multirow{2}{*}{-} & \multicolumn{2}{c|}{Height} & \multicolumn{2}{c|}{Age} \\ \cline{3-6} 
 &  & RMSE & MAE & RMSE & MAE \\ \hline
\multirow{2}{*}{Singh et al.(fusion) \cite{7449696}} & M & \textbf{6.7} & \textbf{5.0} & 7.8 & 5.5 \\ \cline{2-6} 
 & F & 6.1 & 5.0 & 8.9 & 6.5 \\ \hline
\multirow{2}{*}{Kalluri et al. \cite{8683397}} & M & 6.85 & - & 7.60 & - \\ \cline{2-6} 
 & F & 6.29 & - & 8.63 & - \\ \hline
\multirow{2}{*}{Kwasny et al. \cite{kwasny2020joint}} & M & - & - & 7.24 & 5.12 \\ \cline{2-6} 
 & F & - & - & 8.12 & 5.29 \\ \hline
 \multirow{2}{*}{Williams et al. \cite{6639131}} & M & - & 5.37 & - & - \\ \cline{2-6} 
 & F & - & 5.49 & - & - \\ \hline
\multirow{2}{*}{Mporas et al. \cite{mporas2009estimation}} & M & 6.8 & 5.3 & - & - \\ \cline{2-6} 
 & F & 6.3 & 5.1 & - & - \\ \hline
 \multirow{2}{*}{\textbf{[Ours]} baseline} & M & 7.3 & 5.6 & 7.7 & 5.5 \\ \cline{2-6} 
 & F & 6.3 & 5.2 & 8.2 & 6.0 \\ \hline
\multirow{2}{*}{\textbf{[Ours]} ST - Height} & M & 8.1 & 5.9 & - & - \\ \cline{2-6} 
 & F & \textbf{6.0} & \textbf{4.9} & - & - \\ \hline
\multirow{2}{*}{\textbf{[Ours]} ST - Age} & M & - & - & 6.96 & 4.8 \\ \cline{2-6} 
 & F & - & - & 7.6 & 5.1 \\ \hline
\multirow{2}{*}{\begin{tabular}[c]{@{}c@{}}\textbf{[Ours]} MT
\end{tabular}} & M & 7.5 & 5.8 & \textbf{6.8} & \textbf{4.8} \\ \cline{2-6} 
 & F & 6.5 & 5.1 & \textbf{7.4} & \textbf{5.0} \\ \hline
\end{tabular}
\caption{Results comparison of height and age on TIMIT test set. MT=Multi Task, ST=Single Task, M=Male, F=Female}
\label{table:1}
\end{table}

\vspace{-2em}

\begin{table}[h!]
\centering
\begin{tabular}{|c|c|}
\hline
Method & \begin{tabular}[c]{@{}c@{}}Gender \\ Accuracy\end{tabular} \\ \hline
Yao et al. \cite{9002874} & 98.0 \\ \hline
Kwasny et al. \cite{kwasny2020joint} & \textbf{99.6} \\ \hline
\textbf{[Ours]} baseline & 99.2 \\ \hline
\textbf{[Ours]} ST & 99.2 \\ \hline
\textbf{[Ours]} MT &  99.1\\ \hline
\end{tabular}
\caption{Results comparison of Gender Accuracy on TIMIT test set. MT=Multi Task, ST=Single Task}
\label{table:2}
\end{table}

\vspace{-2em}

\begin{table}[h!]
\centering
\begin{tabular}{|c|c|c|c|c|c|}
\hline
\multirow{2}{*}{Method} & \multirow{2}{*}{-} & \multicolumn{2}{c|}{Height} & \multicolumn{2}{c|}{Age} \\ \cline{3-6} 
 &  & RMSE & MAE & RMSE & MAE \\ \hline
\multirow{2}{*}{Kalluri et al. \cite{kalluri2020nisp}}  & M & \textbf{6.13} & \textbf{5.16} & 5.63 & 3.80 \\ \cline{2-6} 
 & F & 6.70 & 5.30 & \textbf{4.99} & \textbf{3.55} \\ \hline
\multirow{2}{*}{\textbf{[Ours]} MT} & M & 6.49 & 5.43 & \textbf{5.55} & \textbf{3.80} \\ \cline{2-6} 
 & F & \textbf{6.44} & \textbf{5.17} & 6.25 & 4.38 \\ \hline
\end{tabular}
\caption{Results comparison of height and age on NISP test set. MT=Multi Task, M=Male, F=Female}
\label{table:3}
\end{table}

\vspace{-2em}

\section{Conclusions}
This paper proposed a semi-supervised learning method to improve the estimation of height, age, and gender of a speaker from the raw audio signal in a multi-task setting. The supervised speaker profiling task is combined with an unsupervised speaker presentation learning task and learns a representation that is rich in speaker information and helps speaker profiling task. The experiments show good performance in estimating the height, age, and gender of a speaker better than the earlier methods. As the semi-supervised learning method learns the speaker representations, these representations can also be used for speaker-specific tasks like speaker recognition/verification, the age, height, and gender information may help to distinguish between speakers better. 



\bibliographystyle{IEEEtran}

\bibliography{mybib}


\end{document}